\documentclass[12pt]{iopart}

\usepackage{iopams}
\usepackage[pdftex]{graphicx}
\usepackage{hyperref}

\usepackage{diagbox}

\newtheorem{lemma}{Lemma}
\newtheorem{theorem}{Theorem}
\newtheorem{remark}{Remark}

\newcommand{\Hess}{{\rm Hess}}
\newcommand{\ba}{{\bf a}}
\newcommand{\J}{J}
\newcommand{\sinc}{{\rm sinc}\,}

\begin{document}

\title[Quantum control landscape for ultrafast generation of single-qubit 
...]{Quantum control landscape for ultrafast generation of single-qubit 
phase shift quantum gates\footnote{Accepted for publication in {\it Journal of 
Physics A: Mathematical and Theoretical} on April 1, 2021. This Accepted Manuscript is available for reuse under a CC BY-NC-ND licence after the 12 
month embargo period provided that all the terms of the licence are adhered to.}}

\author{Boris O Volkov$^{1}$, Oleg V Morzhin$^1$ and Alexander N Pechen$^{1,2}$
}
\address{$^1$ Department of Mathematical Methods for Quantum Technologies, Steklov Mathematical Institute of Russian Academy of Sciences, 8 Gubkina Str., Moscow, 119991, Russia; \url{www.mi-ras.ru/eng/dep51}}
\address{$^2$ National University of Science and Technology "MISIS", 4 Leninsky Prosp., Moscow 119991, Russia}

\eads{
\mailto{borisvolkov1986@gmail.com}, 
\mailto{morzhin.oleg@yandex.ru},\\
\mailto{Corresponding author: apechen@gmail.com}; \url{www.mathnet.ru/eng/person17991}} 

\begin{abstract} 
In this work, we consider the problem of ultrafast controlled generation 
of single-qubit phase shift quantum gates. Globally optimal control is a control which realizes the gate with maximal possible fidelity. Trap is a 
control which is optimal only locally but not globally. It was shown before that traps do not exist for controlled generation of arbitrary single-qubit quantum gates for sufficiently long times, as well as for fast control of quantum gates other than phase shift gates. Ultrafast generation of phase-shift gates was missed in the previous analysis. In this work we show, combining analytical and numerical optimization methods such as Gradient Ascent Pulse Engineering (GRAPE), differential evolution, and dual annealing, that control landscape for ultrafast generation of phase shift 
gates is also free of traps.  
Mathematical analysis of quantum control landscapes, which aims to prove either absence or existence of traps for quantum control objective functionals, is an important topic in quantum control. In this work, we provide 
a rigorous analysis of quantum control landscapes for ultrafast generation of single-qubit quantum gates and show, combining analytical methods based on a sophisticated analysis of spectrum of the Hessian, and numerical 
optimization methods such as Gradient Ascent Pulse Engineering (GRAPE), differential evolution, and dual annealing, that control landscape for ultrafast generation of phase shift gates is free of traps.  
\end{abstract}

\noindent{\it Keywords\/}: quantum control, control landscape, qubit, phase shift gate

\section{Introduction}  

Control of atomic and molecular systems is an important branch of modern science with various existing and prospective applications in various directions of quantum technologies~\cite{Shapiro_Brumer_book,RiceBook,TannorBook,LetokhovBook,AlessandroBook,Moore2011,Brif2012,Lidar2010,Glaser2015Report}. In application to quantum computing, it can be used to generate with high fidelity quantum gates in a minimal time~\cite{Google}. 

In this work we consider the problem of generating single qubit phase shift quantum gates. Single qubit phase shift gate with phase  $\phi\in[0,2\pi)$ in the computational basis can be represented as a unitary $2\times 2$ matrix
\[
R_\phi=\left(\begin{array}{cc}
1 & 0\\
0 & e^{i\phi}
\end{array}
\right).
\]
Up to an unphysical phase factor one has $R_\phi=W_{\phi/2}$, where $W_\phi=e^{i\phi\sigma_z}$ and $\sigma_z$ is the $z$-Pauli matrix. On ultrafast time scale, the influence of the environment often can be negligible and the dynamics of the qubit can be approximately described by Schr\"odinger equation:
\begin{equation}
\label{Shred}
i\frac{dU^f_t}{dt}=(H_0+f(t)V)U^f_t,\qquad U^f_{t=0}=\mathbb I.
\end{equation}
Here  $H_0$ and $V$ are the free and interaction Hamiltonians (Hermitian $2\times 2$-matrices), and $f$ is  a coherent control. The free and interaction Hamiltonians are assumed to be non-commutative, $[H_0,V]\neq 0$, to exclude trivial case. 

Typical choices for the space of controls are the spaces $L^1=L^1([0,T],\mathbb{R})$ and $L^2=L^2([0,T],\mathbb{R})$ which consist of all Lebesgue measurable functions $f:[0,T]\to\mathbb R$ such that, correspondingly, $\int\limits_0^T |f(t)|dt <\infty$ and $\int\limits_0^T f^2(t)dt <\infty$ (strictly speaking, elements of these spaces are equivalence classes of functions which coincide almost everywhere). The latter is a subspace of the former,  $L^2\subset L^1$. The space $L^1$ is the most general space of controls for which the Schr\"odinger equation~(\ref{Shred}) by Carath\' eodory's existence theorem for every control $f$ has a unique absolutely continuous solution. 
The space $L^2$ is a Hilbert space (while $L^1$ is not), and this property is convenient for Hessian analysis performed in this work. For this reason, we use $L^2$ in this work as the control space. For numerical optimization we use finite dimensional subspaces of piecewise constant controls 
defined explicitly in Sec.~\ref{subsec5.2}.

The goal of quantum control is to find a control $f$ such that the induced evolution $U_T^f$ is as close as possible to the target gate $W\in SU(2)$. The problem of generating a single qubit gate $W$ can be formulated as problem of maximizing the objective functional,
\begin{equation}\label{functional}
J_W[f]=\frac14|\Tr(W^\dagger U^f_T)|^2\to\max.
\end{equation}
Indeed, the maximum value $J_W[f_*]=1$ is attained if and only if the control $f_*$  is such that $U_T^{f_*}=e^{i\alpha}W$ for some (physically unrelevant) phase $\alpha$.

Trap is a control which is optimal only locally but not globally. The analysis of traps is important for practical design of control fields, where 
it can help to select between global and local search optimal control protocols. Global search protocols (e.g., genetic algorithms, simulated annealing, stochastic optimization) are the better choice if traps do exist, while local search protocols (e.g., gradient-type) are preferable if it is theoretically known that a given quantum control objective has no traps. In this situation local search protocols could be more efficient for finding optimal controls, while in the presence of traps their work can be hindered at a local maximum thereby preventing arrival of the maximum of the objective functional. Choosing an efficient strategy is especially important for laboratory optimization, where high number of iterations of 
the protocol is a time/energy/cost expensive.

An important problem in quantum control is to investigate whether traps can or can not exist in quantum control landscapes~\cite{RHR,HR,Chakrabarty2007,Pechen2008,Wu2008,Pechen2011,Pechen2012,PechenIl'in2014,FouquieresSchirmer,Larocca2018}. The problem in its completeness is not yet solved~\cite{Zhdanov2018,Russell2018}. Much effort is directed towards ultrafast control at quantum speed limit~\cite{Caneva2009, Pechen2012, Bason2012, Hegerfeldt2013, Avinadav_Fischer_London_Gershoni_article_2014, Hegerfeldt2014, PechenIl'in2017, Mortensen2018, Lin_Sels_Wang_article_2020,LamPRX2021}. For this case, the absence of traps for control landscape of a two-level Landau-Zener system at times around or greater quantum speed limit was rigorously proved in~\cite{Pechen2012}, and for some cases below quantum speed limit in~\cite{PechenIl'in2017, PechenIl'in2016} (see also~\cite{PechenIl'in2018}). Numerical investigation of the control landscape for this system for a different from the present work control objective (transition probability) for times around the quantum speed limit was performed 
in~\cite{Larocca2018}. Severe restrictions, e.g. considering piecewise constant controls with a small number of constant components, may destroy the trap-free property as shown for example on figure~1 in Ref.~\cite{Pechen2012}. Thus when we discuss the absence of traps, we mean the restriction-free case with controls belonging to the functional space.

In this work we show, combining analytical and numerical results, that ultrafast generation of phase shift gates is also free of traps. The structure of the work is the following. In Sec.~\ref{Sec:2} we outline known results. In Sec.~\ref{Sec:3} main theoretical result of this work is formulated. Sec.~\ref{Sec:4} contains formulations and proofs of several lemmas 
which when combined give proof of the main theorem. Sec.~\ref{Sec:5} contains numerical analysis of the control landscape for various values of the parameters $\phi$ and $T$ using Gradient Ascent Pulse Engineering (GRAPE)~\cite{GRAPE} 
stochastic zeroth-order optimization methods of differential evolution~\cite{Differential_Evolution_SciPy, Storn_Price_article_1997}~(DE) and dual 
annealing~\cite{Dual_Annealing_SciPy, Tsallis_Stariolo_article_1996, Xiang_Gong_article_2000}~(DA).

\section{Previous results}\label{Sec:2}

In~\cite{Pechen2012,PechenIl'in2014} it was proved that if $T$ is large enough then the control objective $J_W[f]$ for a qubit has no traps. Later 
in~\cite{PechenIl'in2016} for arbitrarily small time and for any $W$ such 
that  $W\ne e^{i\varphi_W\sigma_z}$ (i.e., for any target gate except of phase shift gates) it was proved that traps also do not exist. For phase-shift gates traps were shown to not exist for any $T>0$ if $\varphi_W\in (0,\pi/2)$ and  for any $T>\pi-\varphi_W$ if $\varphi_W\in [\frac \pi 2,\pi]$, in the units when the Hamiltonian is properly normalized. The only absent up to now case has been the analysis of traps for generating phase-shift gates with $\varphi_W\in [\frac \pi 2,\pi]$ at a small time.

To explicitly formulate these previously known results on the absence of traps for controlled single qubit gate generation, consider the special constant control $f(t)=f_0$ and time $T_0$:
\begin{eqnarray}
f_0&:=&\frac{-\Tr H_0\Tr V+2\Tr(H_0V)}{(\Tr V^2)^2-2\Tr(V^2)},\\
T_0&:=&\frac {\pi}{\|H_0-\mathbb I \Tr H_0/2+f_0(V-\mathbb I  \Tr V/2)\|}.
\end{eqnarray}

The following result was proved in~\cite{PechenIl'in2014,PechenIl'in2016}. 

\begin{theorem}\label{theorem2016-1}
Let $W\in SU(2)$ be a single qubit quantum gate.
If $[W,H_0+f_0V]\neq0$ then for any $T>0$ traps do not exist. If $[W,H_0+f_0V]=0$ then any control, except possibly $f\equiv f_0$, is not trap for any $T>0$ and the control  $f_0$ is not  trap for $T>T_0$.
\end{theorem}

The case of whether control $f_0$ can be trap for $T\leq T_0$ was partially studied in~\cite{PechenIl'in2016}. Without loss of generality it is sufficient to consider the case $H_0=\sigma_z$ and $V=v_x\sigma_x+v_y\sigma_y$, where $\upsilon_x,\upsilon_y\in \mathbb{R}$ ($\upsilon_x^2+\upsilon_y^2>0$) and $\sigma_x,\sigma_y,\sigma_z$ are the Pauli matrices:
\begin{equation}
\sigma_x=\Bigg(\begin{array}{*{20}{c}}0 & 1 \\ 1 & 0\end{array}\Bigg), \qquad  \sigma_y=\Bigg(\begin{array}{*{20}{c}}0 & -i \\ i & 0\end{array}\Bigg), \qquad \sigma_z=\Bigg(\begin{array}{*{20}{c}} 1 & 0 \\ 0 & -1\end{array}\Bigg).
\end{equation}
In this case, the special time is $T_0=\frac {\pi}2$ and the special control is $f_0=0$.

By theorem~\ref{theorem2016-1}, if $[W,\sigma_z]\neq 0$, then  for any  $T>0$ there are no traps for $J_W$. If $[W,\sigma_z]=0$, then $W=e^{i\varphi_W \sigma_z+i\beta}$, where  $\varphi_W\in (0,\pi]$ and $\beta\in [0,2\pi)$. The phase can be neglected, so without loss of generality we set $\beta=0$. Below we consider only such gates. The following result was proved in~\cite{PechenIl'in2016}. 

\begin{theorem}\label{theorem2016-2}
Let $W=e^{i\varphi_W \sigma_z}$. If $\varphi_W\in (0,\frac{\pi}{2})$, then for any $T>0$ there are no traps. If $\varphi_W\in [\frac \pi 2,\pi]$, then for any $T>\pi-\varphi_W$ there are no traps.
\end{theorem}

For fixed $\varphi_W$ and $T$ the value of the objective evaluated at $f_0$ is
\begin{equation}\label{JD2}
J_W[f_0]=\cos^2{(\varphi_W+T)}.
\end{equation}
If $\varphi_W+T=\pi$ then $J_W[f_0]=1$ and $f_0$ is a global maximum. 

If $\varphi_W+T=\frac{\pi}2$ and $\varphi_W+T=\frac{3\pi}2$ then $J_W[f_0]=0$ and $f_0$ is a global minimum. 

In the present work we study the remaining case when $\varphi_W\in [\frac 
\pi 2,\pi]$ and $T<\pi-\varphi_W$. In this case $J_W[f_0]<1$. We prove that for this case the special control $f_{0}= 0$ is a critical point  for the objective functional $J_W$ and that this point is not a saddle point. We prove that the Hessian of $J_W$ at $f_0$ is an injective compact operator which has only negative eigenvalues. In this case $f_0$ could be either a (1) global maximum, (2) trap, or (3) trap in a more weak sense, such that a restriction of $J_W$ to any finite dimensional subspace of $L^2([0,T],\mathbb{R})$ would have local maximum at $f_0$ while $f_0$ is not 
a point of  global maximum. Performed numerical simulations show that the 
first case has place, i.e., $f_0$ is a point of global maximum, while giving a rigorous proof of this finding remains an open problem. 
The numerical results also show that for $\frac{\pi}{2} \leq \varphi_W \leq \pi$ and $0 < T \leq \frac{\pi}{2}$ achieving the objective functional 
value~1, i.e., providing exact generation of phase shift gate, requires minimal time $T_{\min} = \pi-\varphi_W$.  

\section{Main theorem}\label{Sec:3}
We use the notations  $Y=W^\dagger U^f_T$ and $V_t=U^{f\dagger}_t VU^f_t$ as considered in Ref.~\cite{PechenIl'in2016}. 

The Taylor expansion of the functional $J_W$ at $f$ up to the second order has the form:
\begin{eqnarray}
\fl J_W[f+\delta f] &=& J_W[f]+\int_0^T\frac{\delta J_W}{\delta f(t)}\delta f(t)dt\nonumber \\
\fl &&+\frac 12 \int_0^T\int_0^T \Hess(t,s)\delta f(t)\delta f(s)dtds+o(\|\delta f\|^2_{L^2}), \quad \delta f\to 0.
\end{eqnarray}
The linear term is determined by the integral kernel of the Fr\'echet derivative, 
\[
\frac{\delta J_W}{\delta f(t)}=\frac 12 \Im(\Tr Y^\ast \Tr(YV_t))
\]
and determines the gradient of the objective; the second order term is the integral kernel of the Hessian,
\[
\Hess(t,s)=
\left\{\begin{array}{l}
\frac 12 \Re(\Tr(YV_{t})\Tr(Y^\ast V_{s})-\Tr(YV_{s}V_{t})\Tr Y^\ast)
,\textrm{ if } s\geq t\\
\frac 12 \Re(\Tr(YV_{s})\Tr(Y^\ast V_{t})-\Tr(YV_{t}V_{s})\Tr Y^\ast)
, \textrm{ if } s<t.
\end{array}\right.
\]
The control $f_0=0$ is a critical point, i.e., gradient of the objective evaluated at this control is zero. The Hessian at $f_0=0$ has the form (see~\cite{PechenIl'in2016}):
\begin{equation}\label{Hess}
\Hess(s,t)=-2\upsilon^2\cos{\varphi}\cos{(2|t-s|+\varphi)},
\end{equation}
where $\varphi=-\varphi_W-T$ and $\upsilon=\sqrt{\upsilon^2_x+\upsilon^2_y}$.

Let us consider the following cases:
\begin{itemize}
\item $(\varphi_W,T)$ belongs to the triangle domain
\[
\hspace*{-1.2cm}
\mathcal{D}_1 := \left\{ (\varphi_W, T)~:~ 0<T<\frac{\pi}{2}, \quad \frac{\pi}{2} \leq \varphi_W < \pi - T \right\}; 
\] 
\item $(\varphi_W,T)$ belongs to the set
\[
\hspace*{-1.2cm}
\mathcal{D}_2 := \left\{ (\varphi_W, T)~:~ 0 < T \leq \frac{\pi}{2}, \quad \varphi_W = \pi - T \right\};
\]
\item $(\varphi_W,T)$ belongs to the triangle domain
\[
\hspace*{-1.2cm}
\mathcal{D}_3 := \left\{ (\varphi_W, T)~:~ 0 < T \leq \frac{\pi}{2}, \quad \pi - T < \varphi_W \leq \pi, \quad  (\varphi_W,T)\neq (\pi, \frac {\pi}2) \right\}.
\]
\item $(\varphi_W,T)$ belongs to the square domain without the diagonal
\[
\hspace*{-1.2cm}
\mathcal{D}_4 := \left\{(\varphi_W, T)~:~ 0 < T \leq \frac{\pi}{2}, \quad 0 < \varphi_W <\frac {\pi}2 , \varphi_W+T\neq\frac{\pi}2\right\}.
\]
\end{itemize} 

Our main result is the following theorem. 
\begin{theorem}
If $(\varphi_W,T)\in \mathcal{D}_1\cup \mathcal{D}_3\cup \mathcal{D}_4$ then the Hessian of the objective functional $J_W$ at $f_0=0$ is an injective compact  operator on $L^2([0,T],\mathbb{R})$. Moreover,
\begin{enumerate}
\item If $(\varphi_W,T)\in \mathcal{D}_1$, then Hessian at $f_0$ has only 
negative eigenvalues.
\item If $(\varphi_W,T)\in \mathcal{D}_3\cup \mathcal{D}_4$  then Hessian 
at $f_0$ has both negative and positive eigenvalues. In this case, the special control $f_0=0$ is a saddle point for the objective functional. 
\end{enumerate}
\end{theorem}
\begin{remark}
The second case was previously proved by Pechen and Il'in in~\cite{PechenIl'in2016}  using a different method. The fist case has not been previously considered and is a new result of this work.
\end{remark}

\begin{remark}
As is mentioned in the introduction, in the previous works~\cite{PechenIl'in2014,PechenIl'in2016} the case when the control belongs to the  space $L^1([0,T],\mathbb{R})$ was considered. In this case the Schr\"odinger equation~(\ref{Shred}) by Carath\' eodory's existence theorem for every control $f$ has a unique absolutely continuous solution.

Let $\{e_n\}$ be an orthonormal basis in the Hilbert space $L^2([0,T],\mathbb{R})$, which consists of the eigenvectors 
of the operator $\Hess$ and $\{\lambda_n\}$ be the corresponding  eigenvalues. Any $\delta f \in  L^2([0,T],\mathbb{R})$
can be expanded in the Fourier series $\delta f=\sum_{i=1}^{\infty}c_ne_n$. Then for the quadratic form generated by $\Hess$ we have the expression 
\[
(\Hess \,\delta f,\delta f)=\int_0^T\int_0^T \Hess(t,s)\delta f(t)\delta f(s)dtds=\sum_{i=1}^{\infty}\lambda_ic_i^2.
\]
If all eigenvalues  $\{\lambda_n\}$ are negative, then the quadratic form 
 is  strictly negative, i.e. $(\Hess \,\delta f,\delta f)<0$ for any $\delta f\neq 0$. In this case, $f_0$ is at least a strict local maximum on any finite dimensional subspace of $L^2([0,T],\mathbb{R})$. In difference to~\cite{PechenIl'in2014,PechenIl'in2016}, where controls $f\in L^1([0,T],\mathbb{R})$ were considered, in the present paper we consider the case $f\in L^2([0,T],\mathbb{R})$, because in general the vectors $\{e_n\}$ do 
not form the Schauder basis in the space $L^1([0,T],\mathbb{R})$.
\end{remark}

\begin{remark}
A class of $n$-level systems  with $n\ge 4$ which have traps was discovered in~\cite{FouquieresSchirmer}. These traps are constant controls. The  Hessian of the objective functional computed at these controls has the form
\begin{eqnarray}
\label{trigHess}
Hess(t,s)=A\cos(\varepsilon|t-s|)+B\sin(\varepsilon|t-s|)+C.
\end{eqnarray}
It was shown in~\cite{FouquieresSchirmer} that if $A$, $B$, and $C$ satisfy certain relations then the Hessian~(\ref{trigHess}) is strictly sign-definite. For the Hessian~(\ref{Hess}) these relations do not hold and we can not use the method of~\cite{FouquieresSchirmer} for analyzing this case. For this reason in the present paper we investigate the spectrum and eigenvalues of Hessian~(\ref{Hess}) to access properties of the control landscape.
\end{remark}

\section{Proof of the main theorem}\label{Sec:4}
In this section we will prove Theorem 3. First we will prove that  if $(\varphi_W,T)\in \mathcal{D}_1\cup \mathcal{D}_3\cup \mathcal{D}_4$ then 
the Hessian of the objective functional $J_W$ at $f_0=0$ is an injective compact  operator on $L^2([0,T],\mathbb{R})$. 

If $(\varphi_W,T)\in \mathcal{D}_1\cup \mathcal{D}_3\cup \mathcal{D}_4$, then $\sin2\varphi=-\sin2(\varphi_W+T)\neq 0$.
Instead of Hessian, we can consider the integral operator $K=\frac 1{\upsilon^2\sin2\varphi} \Hess $ with integral kernel:
\begin{equation}
K(s,t)=-\frac{\cos{(2|t-s|+\varphi)}}{\sin{\varphi}},
\end{equation}
Let
\begin{eqnarray*}
\fl h(t)=(K g)(t)&=&-\frac 1{\sin{\varphi}}\int_0^T\cos{(2|t-s|+\varphi)}g(s)ds\\
\fl &=&-\frac 1{\sin{\varphi}}\int_0^t\cos{(2t-2s+\varphi)}g(s)ds\\
\fl &&-\frac 1{\sin{\varphi}}\int_t^T\cos{(2s-2t+\varphi)}g(s)ds.
\end{eqnarray*}
Assume that the function $g$ is continuous.
Then  $h=Kg$ is  $C^2$-smooth function on $[0,T]$.
The first and second derivatives of  $h$ have the form
\begin{eqnarray}
\fl h'(t)=\frac 2{\sin{\varphi}}\left(\int_0^t\sin{(2t-2s+\varphi)}g(s)ds-\int_t^T\sin{(2s-2t+\varphi)}g(s)ds\right),\\
\fl h''(t)=\frac 4{\sin{\varphi}}\left(\int_0^t\cos{(2t-2s+\varphi)}g(s)ds+\int_t^T\cos{(2s-2t+\varphi)}g(s)ds\right)\nonumber\\
+4g(t)=-4h(t)+4g(t).
\end{eqnarray}
So if  $g$ is continuous and  $h=Kg$, then the following equality holds
\begin{equation}\label{eq1!}
h''(t)+4h(t)=4g(t).
\end{equation}

Let us show that the equality~(\ref{eq1!}) holds in the weak sense for an 
arbitrary $g\in L^2([0,T],\mathbb{R})$ and  $h=Kg$.
Consider an arbitrary  test function $f$, i.e. $f\in C^\infty([0,T],\mathbb{R})$ and  the support of $f$
belongs to the open interval $(0,T)$.
Fubini's theorem implies that
\begin{equation}\label{Fub!!!}
\fl \langle f'',h \rangle=\int_0^Tf''(t) (Kg)(t)dt
=-\frac 1{\sin{\varphi}}\int_0^T\left(\int_0^Tf''(t)\cos{(2|t-s|+\varphi)} dt\right)g(s)ds.
\end{equation}
The internal integral can be integrated two times by parts:
\begin{equation}
\fl \int_0^Tf''(t)\cos{(2|t-s|+\varphi)}dt
=-4\int_0^Tf(t)\cos{(2|t-s|+\varphi)}dt-4\sin{\varphi}f(s).
\end{equation}
Substituting this expression into~(\ref{Fub!!!}), we obtain
\begin{eqnarray}\label{weak}
\fl \langle f,h''\rangle=\langle f'',h \rangle&=&\frac 4{\sin{\varphi}}\int_0^T\left(\int_0^Tf(t)\cos{(2|t-s|+\varphi)}dt+4\sin{\varphi}f(s)\right)g(s)ds\nonumber\\
&=&-4\langle f,h \rangle+4\langle f,g\rangle.
\end{eqnarray}
Thus equality~(\ref{eq1!}) holds in the weak sense.
As a result, we get that if $h=0$, then $g=0$. Hence  the operator  $K$ is an injective compact operator on $L^2([0,T],\mathbb{R})$.  

For any continuous  $g$, we can find  $h=Kg$ as a unique solution of ODE~(\ref{eq1!}), which satisfies the initial conditions
\begin{eqnarray}
h(0)&=&-\frac 1{\sin{\varphi}}\int_0^T\cos{(2s+\varphi)}g(s)ds, \label{bound1}\\
h'(0)&=&-\frac {2}{\sin{\varphi}}\int_0^T\sin{(2s+\varphi)}g(s)ds. \label{bound2}
\end{eqnarray}

Let $\mu$ be an eigenvalue of the operator $K$ and $g$ be the corresponding eigenfunction, so that $h=Kg=\mu g$. 
Let $\lambda=1/\mu$.
Then using~(\ref{eq1!}), we obtain that
\begin{equation}\label{hh}
h''(t)=4(\lambda-1)h(t).
\end{equation}
In the following sections, we examine whether eigenvalue $\mu$  can be positive.

\subsection{Case $\lambda<1$}
Consider the case $\lambda<1$.
Let $a^2=4(1-\lambda)$ and $a>0$. If $h$ satisfies~(\ref{hh}) then $h$ has the form $h(t)=b\cos at+c\sin at$ and  
\[
g(t)=\frac{4-a^2}4(b\cos at+c\sin at).
\]

Substituting  $h$ and  $g$ in the initial condition~(\ref{bound1}), we obtain
\begin{eqnarray}
\label{bound11}
\fl h(0)=b&=&-\frac 1{\sin{\varphi}}\int_0^T\cos{(2s+\varphi)}g(s)ds\nonumber\\
\fl &=&\frac {a^2-4}{4\sin{\varphi}}b\int_0^T\cos{(2t+\varphi)}\cos atdt +\frac{a^2-4}{4\sin{\varphi}}c\int_0^T\cos{(2t+\varphi)}\sin atdt.
\end{eqnarray}
This equality can been rewritten in the form
\begin{equation}\label{lless1eq1}
0=A_{11}(a)b+A_{12}(a)c,
\end{equation}
where
\begin{eqnarray*}
A_{11}(a)&=&2\sin{\varphi}-a\sin{(aT)}\cos{(2T+\varphi)}+2\cos{(aT)}\sin{(2T+\varphi)},\\
A_{12}(a)&=&a\cos{(aT)}\cos{(2T+\varphi)}+2\sin{(aT)}\sin{(2T+\varphi)}-a\cos{\varphi}.
\end{eqnarray*}

Substituting $h$ and  $g$ in the initial conditions~(\ref{bound2}), we obtain
\begin{eqnarray}\label{bound22}
\fl h'(0) &=& ac = -\frac 2{\sin{\varphi}}\int_0^T\sin{(2s+\varphi)}g(s)ds 
\nonumber \\
\fl &=& \frac {a^2-4}{2\sin{\varphi}}b\int_0^T\sin{(2t+\varphi)}\cos(at)dt + \frac{a^2-4}{4\sin{\varphi}}c\int_0^T\sin{(2t+\varphi)}\sin(at)dt. 
\end{eqnarray}
This equality can been rewritten in the form
\begin{equation}
\label{lless1eq2}
0=A_{21}(a)b+A_{22}(a)c,
\end{equation}
where
\begin{eqnarray*}
A_{21}(a)
&=&a\sin{(aT)}\sin{(2T+\varphi)}+2\cos{(aT)}\cos{(2T+\varphi)}-2\cos{\varphi}, \\
A_{22}(a)
&=&-a\sin{\varphi}+2\sin{(aT)}\cos{(2T+\varphi)}-a\cos{(aT)}\sin{(2T+\varphi)}.
\end{eqnarray*}

The function $g(t)=\frac{4-a^2}4(b\cos at+c\sin at)$ is an eigenfunction of the operator $K$ with the eigenvalue $\mu=1/\lambda=4/(4-a^2)$ if and only if $(x_1,x_2)=(b,c)$ is  a nonzero solution of the linear system of equations
\begin{equation}\label{system1}
\left\{\begin{array}{l}
A_{11}(a)x_1+A_{12}(a)x_2=0\\
A_{21}(a)x_1+A_{22}(a)x_2=0.
\end{array}\right.
\end{equation}

Define the function of argument $a\in\mathbb R$
\[
F^1_{\varphi_W,T}(a)=A_{11}(a)A_{22}(a)-A_{12}(a)A_{21}(a).
\]
By direct computations we get
\[
\fl F^1_{\varphi_W,T}(a)=-4a-a^2\sin{(aT)}\sin{(2\varphi_W)}-4\sin{(aT)}\sin{(2\varphi_W)}+4a\cos{(aT)}\cos({2\varphi_W)}.
\]

Nonzero  solutions of the system~(\ref{system1}) exists if and only if $F^1_{\varphi_W,T}(a)=0$. Let us analyze positive roots of the function $F^1_{\varphi_W,T}$.

\begin{lemma}
\label{lemma1}
The function $F^1_{\varphi_W,T}$ has  infinitely many roots in the interval $(2,+\infty)$ for any $(\varphi_W,T)\in \mathcal{D}_1\cup \mathcal{D}_3\cup \mathcal{D}_4$. Hence, operator $K$ has infinitely many negative eigenvalues.
\end{lemma}
\noindent{\bf Proof.}
If $\varphi_W\neq \frac {\pi} 2$ and $\varphi_W\neq \pi$, then let
$a_n={(-\frac{\pi}2+2\pi n)}/T$ and $a'_n={(\frac{\pi}2+2\pi n)}/T$. 
We have
\begin{eqnarray*}
F^1_{\varphi_W,T}(a_n)&\sim& \frac{(2\pi)^2}{T^2}\sin{(2\varphi_W)}n^2, \quad n\to+\infty,\\
F^2_{\varphi_W,T}(a'_n)&\sim&-\frac{(2\pi)^2}{T^2}\sin{(2\varphi_W)}n^2, \quad n\to+\infty.
\end{eqnarray*}
The values of $F^1_{\varphi_W,T}(a_n)$  and $F^1_{\varphi_W,T}(a'_n)$ for 
sufficiently large $n$  have different signs. Since to $F^1_{\varphi_W,T}$ is continuous, for sufficiently large $n$ there  exists at least one root of this function in the interval $(a_n,a'_n)$ and at least one root in 
the interval $(a'_n,a_{n+1})$.

If $\varphi_W= \frac {\pi} 2$,  then $a_n=\frac {\pi+2\pi n}T$ are roots of the function $F^1_{\varphi_W,T}$. If $\varphi_W=\pi$,  then $a_n=\frac {2\pi n}T$ are roots of the function $F^1_{\varphi_W,T}$. This completes the proof.

\begin{lemma}
\label{lemma2}
If $(\varphi_W,T)\in\mathcal {D}_3$ such that $\varphi_W<\pi$, then $F^1_{\varphi_W,T}$ has a root in the interval $(0,2)$. Hence in this case the 
operator $K$ has at least one positive eigenvalue.
\end{lemma}
\noindent{\bf Proof.}
The function  $F^1_{\varphi_W,T}$ can been rewritten as
\[
F^1_{\varphi_W,T}(a)=-4a-(a-2)^2\sin{(aT)}\sin{(2\varphi_W)}+4a\cos{(2\varphi_W+aT)}.
\]
Let $a'=(2\pi-2\varphi_W)/T$. If $(\varphi_W,T)\in\mathcal {D}_3$ such that $\varphi_W<\pi$, then
 we have $a'\in (0,2)$. The following inequality holds
\[
F^1_{\varphi_W,T}(a')=(a'-2)^2\sin^2{(2\varphi_W)}> 0.
\]
If $(\varphi_W,T)\in\mathcal {D}_3$, then $\varphi\in(-\frac 32\pi,-\pi)$.
Hence,
\[
F^1_{\varphi_W,T}(2)=-16\sin^2{\varphi}< 0.
\]
Since the function  $F^1_{\varphi_W,T}$ is continuous, there exists a root  of this function in the interval  $(a',2)$. 
This completes the proof.

\begin{lemma}
\label{lemma3}
If $(\varphi_W,T)\in \mathcal{D}_1$, then the function $F^1_{\varphi_W,T}$ has not roots in the interval $(0,2)$.
\end{lemma}
\noindent{\bf Proof.}
Fix any  $a\in (0,2)$. Let consider the function of two arguments $G_1(\varphi_W,T)=F^1_{\varphi_W,T}(a)$. In other words
\begin{equation*}
\label{Gvarphi}
G_1(\varphi_W,T)=-4a-(a^2+4)\sin{(aT)}\sin{(2\varphi_W)}+4a\cos{(aT)}\cos{(2\varphi_W)}.
\end{equation*}
Let consider the set  ${\rm int}\,\mathcal{D}_1$ of internal points of $\mathcal{D}_1$, i.e.
\[
{\rm int}\, \mathcal{D}_1=\{(\varphi_W,T)\colon \varphi_W\in (\pi/2,\pi),T>0,\varphi_W+T<\pi\}.
\]
Let $\overline{\mathcal{D}_1}$ and $\partial \mathcal{D}_1$ denote the closure and the boundary of the set $\mathcal{D}_1$. We have $\partial \mathcal{D}_1=I_1\cup I_2 \cup \mathcal D_2$, where $I_1=\{(\pi/2,T)\colon T\in [0,\pi/2]\}$, $I_2=\{(\varphi_W,0)\colon \varphi_W\in [\pi/2,\pi]\}$.
 
Now we  will prove that maximum value of $G_1$ is zero on the boundary $\partial\mathcal{D}_1$. Consider the cases:
\begin{enumerate}  
\item If $(\varphi_W,T)\in I_1$ then 
\[
G_1(\varphi_W,T)=G\left(\frac \pi 2,T\right)=-4a-4a\cos{(aT)} <0.
\]
\item 
If $(\varphi_W,T)\in I_2$ then 
\[
G_1(\varphi_W,T)=G_1(\varphi_W,0)=-4a+4a\cos{(2\varphi_W)}\leq 0.
 \]
Note that  we have equality only if $(\varphi_W,T)=(\pi,0)$.
\item
Now let us show that  $G_1(\varphi_W,T)<0$ for all $(\varphi_W,T)\in \mathcal D_2$.
For this purpose we consider the function $H_1(x)=G_1(\pi-x,x)$ for $x\in [0,\frac 12 \pi]$, 
\[
H_1(x)=-4a+(a^2+4)\sin{(ax)}\sin{(2x)}+4a\cos{(a x)} \cos{(2 x)}.
\]
The derivative of  $H_1$ is
\begin{equation*}
H_1'(x)=\frac 1x(a^2 - 4)\sin{(2x)}\sin{(ax)}(ax \cot{(ax)} - 2x\cot{(2 
x)}).
\end{equation*}
The function  $f(x)=x\cot{x}$ strictly decreases in the interval $(0,\pi)$ and, hence, $ax\cot{(ax)}-2x\cot{(2x)}>0$ and  $H_1'(x)<0$  for all $x\in(0,\frac \pi 2)$. Thus the  function $H_1$ strictly decreases in the closed interval $[0,\frac \pi 2]$. Since $H_1(0)=0$, we have
$H_1(x)<0$ for all $x\in(0,\frac \pi 2]$. It implies that $G_1(\varphi_W,T)<0$ for any $(\varphi_W,T)\in \mathcal{D}_2$. 
\end{enumerate}  

Now we will show that the function $G_1$ has not critical points on ${\rm 
int}\mathcal{D}_1$. Assume that there exists a critical point
$(\varphi_W^\ast,T^\ast)\in {\rm int}\mathcal{D}_1$.
Then 
\begin{equation}\label{criticalsystem1}
\fl \left\{\begin{array}{l}
\partial_{\varphi_W}G_1(\varphi_W^\ast,T^\ast)=-2(a^2+4)\sin{(aT^\ast)}\cos{(2\varphi_W^\ast)}-8a\cos{(aT^\ast)}\sin{(2\varphi_W^\ast)}=0\\
\partial_{T}G_1(\varphi_W^\ast,T^\ast)=-(a^2+4)a\cos{(aT^\ast)}\sin{(2\varphi_W^\ast)}-4a^2\sin{(aT^\ast)}\cos{(2\varphi_W^\ast)}=0.
\end{array}\right.
\end{equation}
Due to $(\varphi_W^\ast,T^\ast)\in {\rm int}\, \mathcal{D}_1$, we have $\sin{(2\varphi_W^\ast)}<0$ and $\sin{(aT^\ast)}>0$. 
Hence, system~(\ref{criticalsystem1}) can been rewritten
as
 \begin{equation}
\label{criticalsystem11}
\left\{\begin{array}{l}
\frac{(a^2+4)}{4a}\cot{(2\varphi_W^\ast)}+\cot{(aT^\ast)}=0\\
\frac{4a}{(a^2+4)}\cot{(2\varphi_W^\ast)}+\cot{(aT^\ast)}=0.
\end{array}\right.
\end{equation}

Due to $a\in (0,2)$, we have  $\frac{4a}{(a^2+4)}\neq \frac{(a^2+4)}{4a}$ 
and, hence,  $\cot{(2\varphi_W^\ast)}=0$ and $\cot{(aT^\ast)}=0$. 
Then  $\varphi_W^\ast=\frac {3\pi}4$. Due to $(\varphi_W^\ast,T^\ast)\in {\rm int}\, \mathcal{D}_1$, we have that
\begin{equation}
0<T^\ast<\pi-\varphi_W^\ast=\frac{\pi}4.
\end{equation}
Hence, $aT^\ast\in (0,\frac {\pi}2)$ and $\cot{(aT^\ast)}>0$. We get a contradiction. The function  $G_1$  has not critical points  on the open set ${\rm int}\,\mathcal{D}_1$. Hence, the function $G_1$ reaches its maximum value on $\overline{\mathcal{D}_1}$ at the boundary point $(\pi,0)$. Due to the point $(\pi,0)$ does not belong to $\mathcal{D}_1$, we have
\begin{equation} 
G_1(\varphi_W,T)<G_1(\pi,0)=0
\end{equation}
for all $(\varphi_W,T)\in \mathcal{D}_1$. Then $a\in (0,2)$ is not a root 
of the function $F^1_{\varphi_W,T}$. This completes the proof.

\subsection{Case $\lambda=1$}
If $\mu=1$ is an eigenvalue of the operator $K$ then
the corresponding eigenfunctions should have the form
\[
h(t)=g(t)=ct+b.
\]

Substituting this function in the initial conditions~(\ref{bound1}), we obtain
\begin{eqnarray}\label{bound11l=1}
\fl h(0)&=&b=-\frac 1{\sin{\varphi}}\int_0^T\cos{(2t+\varphi)}g(t)dt=-\frac 1{\sin{\varphi}}\int_0^T\cos{(2t+\varphi)}(ct+b)dt.
\end{eqnarray}
This equality can been rewritten as
\begin{equation}
\label{l1eq1}
0=B_{11}c+B_{12}b,
\end{equation}
where
\begin{eqnarray}
B_{11}&=&-\cos{\varphi}+\cos{(\varphi+2T)}+2T\sin{(\varphi+2 T)}, \\
B_{12}&=&2(\sin{(2T+\varphi)+\sin{\varphi}}).
\end{eqnarray}
Substituting $h$ and $g$ in the initial conditions~(\ref{bound2}), we obtain
\begin{equation}\label{bound11l=2}
\fl h'(0)=c=-\frac {2}{\sin{\varphi}}\int_0^T\sin{(2t+\varphi)}g(t)dt=-\frac 2{\sin{\varphi}}\int_0^T\sin{(2t+\varphi)}(ct+b)dt.
\end{equation}
This equality can been rewritten as
\begin{equation}\label{l1eq2}
0=B_{21}c+B_{22}b,
\end{equation}
where
\begin{eqnarray*}
B_{21}&=&\sin{\varphi}+\sin{(\varphi+2T)}-2T\cos{(\varphi+2T)},\\
B_{22}&=&2(-\cos{(2T+\varphi)+\cos{\varphi}}).
\end{eqnarray*}

The function $g(t)=ct+b$ is an eigenfunction of the operator $K$ with the eigenvalue $\mu=1$ if and only if $(x_1,x_2)=(c,b)$ is a nonzero solution of the system of linear equations
\begin{equation}\label{system2}
\left\{\begin{array}{l}
B_{11}x_1+B_{12}x_2=0\\
B_{21}x_1+B_{22}x_2=0.
\end{array}\right.
\end{equation}
Such a solution exists if and only if 
\begin{equation}
\Delta=B_{11}B_{22}-B_{12}B_{21}
=-2\sin{\varphi_W}(\sin{\varphi_W}+T\cos{\varphi_W})=0.
\end{equation}

\begin{lemma}\label{lemma4}
If $(\varphi_W,T)\in \mathcal{D}_1$ then $\mu=1$ is not an eigenvalue of the operator $K$.
If $(\varphi_W,T)\in \mathcal{D}_3$  such that $\varphi_W=\pi$ then $\mu=1$ is an eigenvalue of the operator $K$.
\end{lemma}
\noindent{\bf Proof.}
If $x\in (0,\frac {\pi}2)$ then $0<x<\tan x$. Hence, if $(\varphi_W,T)\in 
\mathcal{D}_1$, then 
\[
T<(\pi-\varphi_W)<\tan{(\pi-\varphi_W)}=-\tan{\varphi_W}.
\]
Hence
\begin{equation*}
\Delta=-2\sin{\varphi_W}\cos{\varphi_W}(\tan{\varphi_W}+T)<0.
\end{equation*}
and $\mu=1$ does not belong to  the spectrum  of the operator $K$.

If $\varphi_W=\pi/2$ and $\varphi_W+T<\pi$, then 
\[
\Delta=-2\sin{\varphi_W}(\sin{\varphi_W}+T\cos{\varphi_W})=-2\neq 0.
\]
Hence, $\mu=1$ is not an eigenvalue of $K$. The second statement of the 
lemma is trivial.
This completes the proof.

\subsection{Case $\lambda>1$}

Consider the case $\lambda>1$.
Let $a^2=4(\lambda-1)$ and $a>0$. If $h$ satisfies~(\ref{hh}) then $h$ has the form $h(t)=be^{at}+ce^{-at}$.
Then 
\[
g(t)=\frac{4+a^2}4(be^{at}+ce^{-at}).
\]

Substituting $h$ and $g$ in the initial condition~(\ref{bound1}) gives
\begin{eqnarray*}\label{m11bound11}
\fl h(0)&=&(b+c)=-\frac 1{\sin{\varphi}}\int_0^T\cos{(2s+\varphi)}g(s)ds\\
\fl &=&-\frac {a^2+4}{4\sin{\varphi}}\left(b\int_0^T\cos{(2t+\varphi)}e^{at}dt+c\int_0^T\cos{(2t+\varphi)}e^{-at}dt\right).
\end{eqnarray*}
After calculating the integrals, this equality can be rewritten as
\begin{equation}
\label{lmore1eq1}
0=C_{11}(a)b+C_{12}(a)c,
\end{equation}
where
\begin{eqnarray*}
C_{11}(a)&=&a\cos{\varphi}-2\sin{\varphi}-e^{a T} \left(a \cos{(\varphi+2T)} + 2 \sin{(\varphi+ 2 T)}\right), \\
C_{12}(a)&=&-a \cos{\varphi}-2\sin{\varphi}+e^{-a T} \left(a \cos{(\varphi+2T)} - 2 \sin{(\varphi+ 2 T)}\right).
\end{eqnarray*}

Substituting  $h$ and  $g$ in the initial conditions~(\ref{bound2}) gives
\begin{eqnarray*}\label{mbound11}
\fl h'(0)&=&(ba-ca)=-\frac 2{\sin{\varphi}}\int_0^T\sin{(2s+\varphi)}g(s)ds\\
\fl &=&-\frac {a^2+4}{2\sin{\varphi}}\left(b\int_0^T\sin{(2t+\varphi)}e^{at}dt+c\int_0^T\sin{(2t+\varphi)}e^{-at}dt\right).
\end{eqnarray*}
After calculating the integrals, this equality can been rewritten as
\begin{equation}
\label{lmore1eq2}
0=C_{21}(a)b+C_{22}(a)c,
\end{equation}
where
\begin{eqnarray*}
C_{21}(a)&=&a\sin{\varphi}+2\cos{\varphi}+e^{a T} \left(a \sin{(\varphi+2T)}+2 \cos{(\varphi+ 2 T)}\right), \\
C_{22}(a)&=&-a \sin{\varphi}+2\cos{\varphi}-e^{-a T} \left(2\cos{(\varphi+2T)} + a \sin{(\varphi+ 2 T)}\right).
\end{eqnarray*}

The function $g(t)=\frac{4+a^2}4(be^{at}+ce^{-at})$ is an eigenfunction 
of the operator $K$ with the eigenvalue $\mu=\frac 1{\lambda} =\frac 4{a^2+4}$ if and only if $(x_1,x_2)=(b,c)$ is a nonzero solution of the 
linear system
\begin{equation}\label{system3}
\left\{\begin{array}{l}
C_{11}(a)x_1+C_{12}(a)x_2=0\\
C_{21}(a)x_1+C_{22}(a)x_2=0.
\end{array}\right.
\end{equation}

Let us define the function of the argument $a$,
\[
F^2_{\varphi_W,T}(a)=C_{11}(a)C_{22}(a)-C_{12}(a)C_{21}(a).
\]
Direct calculation gives the expression
\begin{eqnarray}
\label{1111bound11} 
\fl F^2_{\varphi_W,T}(a) &=& 8a-2a^2\sinh{(aT)}\sin{(2\varphi_W)} \nonumber \\
\fl &&-8a\cosh{(aT)}\cos{(2\varphi_W)}
+8\sinh{(aT)}\sin{(2\varphi_W)}.
\end{eqnarray}
Nonzero  solutions of the system~(\ref{system3}) exists if and only if $F^2_{\varphi_W,T}(a)=0$.

\begin{lemma}
\label{lemma5}
If $(\varphi_W,T)\in \mathcal{D}_4$, then $F^2_{\varphi_W,T}$ has at least one positive root. Hence, in this case, the operator $K$ has at least one positive eigenvalue.
\end{lemma}
\noindent{\bf Proof.}
Consider the case  $\varphi_W\in (0,\pi/2)$. Note that
\begin{eqnarray*}
F^2_{\varphi_W,T}(a)&\sim& 8a(1-\cos{(2\varphi_W)}+T\sin{(2\varphi_W)}),\quad a\to 0+\\
F^2_{\varphi_W,T}(a)&\sim& -a^2e^{aT}\sin{(2\varphi_W)},\quad a \to +\infty.
\end{eqnarray*}
If $\varphi_W\in (0,\pi/2)$, then $\sin{(2\varphi_W)}>0$ and hence,
\[
(1-\cos{(2\varphi_W)}+T\sin{(2\varphi_W)})>0.
\]
For sufficiently small $a$ one has $F^2_{\varphi_W,T}(a)>0$, while
for sufficiently large $a$ one has $F^2_{\varphi_W,T}(a)<0$. Since the function $F^2_{\varphi_W,T}$ is continuous, it has at least one positive root. 
This completes the proof.

\begin{lemma}
\label{lemma6}
If $(\varphi_W,T)\in \mathcal{D}_1$, then the function  $F^2_{\varphi_W,T}$ has no positive roots.
\end{lemma}
\noindent{\bf Proof.}
Fix any  $a\in (0,2)$. Consider the function of two arguments $G_2(\varphi_W,T)=F^2_{\varphi_W,T}(a)$,
\begin{equation*}\label{Gvarphi2}
G_2(\varphi_W,T)=8a+2(4-a^2)\sinh{(aT)}\sin{(2\varphi_W)}-8a\cosh{(aT)}\cos{(2\varphi_W)}.
\end{equation*}

We show that the minimal value of $G_2$ is zero on the set $\partial\mathcal{D}_1$ separately for various cases:
\begin{enumerate}  
\item If $(\varphi_W,T)\in I_1$, then 
\[
G_2(\varphi_W,T)=G_2\left(\frac \pi 2,T\right)=8a+8a\cosh{(aT)}> 0.
\]
\item 
If $(\varphi_W,T)\in I_2$, then 
\[
G_2(\varphi_W,T)=G_2(\varphi_W,0)=8a-8a\cos{(2\varphi_W)}\leq 0.
\]
Note that we have the equality only if  $(\varphi_W,T)=(\pi,0)$.
\item
Now let us show that  $G_2(\varphi_W,T)<0$ for all $(\varphi_W,T)\in \mathcal D_2$. For this purpose consider the function $H_2(x)=G_2(\pi-x,x)$, where $x\in [0,\pi/2]$. In other words,
\[
H_2(x)=8a+(2a^2-8)\sinh{(ax)}\sin{(2x)}-8a\cosh{(ax)}\cos{(2x)}.
\]
The derivative of $H_2$ is
\begin{equation*}
H_2'(x)=\frac 2x(4 + a^2)\sin{(2 x)} \sinh{(a x)}(ax \coth{(a x)}-2x\cot{(2x)}).
\end{equation*}
For any  $x\in (0,\frac \pi 2)$ the following estimates hold
\begin{eqnarray*}
2x\cot{(2x)}<1\\
ax\coth{(ax)}>1.
\end{eqnarray*}
This implies that $ax \coth{(a x)}-2x\cot{(2x)}>0$ and  $H_2'(x)>0$ for $x\in (0,\frac \pi 2)$. So the function  $H_2$ strictly increases on $[0,\frac \pi 2]$. Since $H_2(0)=0$, we have $H_2(x)>0$  for any $x\in (0,\frac \pi 2]$. It means that $G_2(\varphi_W,T)>0$ for any $(\varphi_W,T)\in 
\mathcal D_2$.
\end{enumerate}  
 
Now we will show that the function $G_2$ has no critical points on ${\rm int}\mathcal{D}_1$. 
Assume that there exists a critical point
$(\varphi_W^\ast,T^\ast)\in {\rm int}\mathcal{D}_1$.
Then
\begin{equation}\label{criticalsystem2}
\fl
\left\{\begin{array}{l}
\partial_{\varphi_W}G_2(\varphi_W^\ast,T^\ast)=4(4-a^2)\sinh{(aT^\ast)}\cos{(2\varphi_W^\ast)}+16a\cosh{(aT^\ast)}\sin{(2\varphi_W^\ast)}=0\\
\partial_{T}G_2(\varphi_W^\ast,T^\ast)=2a(4-a^2)a\cosh{(aT^\ast)}\sin{(2\varphi_W^\ast)}-8a^2\sinh{(aT^\ast)}\cos{(2\varphi_W^\ast)}=0.
\end{array}\right.
\end{equation}

If $(\varphi_W^\ast,T^\ast)\in {\rm int}\,\mathcal{D}_1$, then $\sin{(2\varphi_W^\ast)}<0$ and $\sinh{(aT^\ast)}>0$. Hence,   system~(\ref{criticalsystem2}) can been rewritten as
\begin{equation}\label{criticalsystem22}
\left\{\begin{array}{l}
\frac{4-a^2}{4a}\cot{(2\varphi_W^\ast)}+\coth{(aT^\ast)}=0\\
\frac{4a}{a^2-4}\cot{(2\varphi_W^\ast)}+\coth{(aT^\ast)}=0.
\end{array}\right.
\end{equation}
Since $\frac{4-a^2}{4a}\neq \frac{4a}{(a^2-4)}$,  we have $\cot{(2\varphi_W^\ast)}=0$ and $\coth{(aT^\ast)}=0$.  But $\coth{(aT^\ast)}>0$. We got a contradiction. 
The function  $G_2$  has not critical points 
on the open set ${\rm int}\,\mathcal{D}_1$. 
Hence, the function $G_2$ reaches its minimum value on $\overline{\mathcal{D}_1}$ at the boundary point $(\pi,0)$. Due to the point $(\pi,0)$ does 
not belong to $\mathcal{D}_1$, we have
\begin{equation} 
G_2(\varphi_W,T)>G_2(\pi,0)=0
\end{equation}
for all $(\varphi_W,T)\in \mathcal{D}_1$. Then $a\in (0,2)$ is not a root 
of the function $F^2_{\varphi_W,T}$.
This completes the proof.

\subsection{Proof of the main theorem.}

In the case $(\varphi_W,T)\in \mathcal{D}_1$, Lemmas~\ref{lemma1},~\ref{lemma3},~\ref{lemma4},~\ref{lemma6} together imply that $K$ has only negative eigenvalues and, hence, $\Hess$ has only negative eigenvalues. If $(\varphi_W,T)\in \mathcal{D}_3$, Lemmas~\ref{lemma1},~\ref{lemma2},~\ref{lemma4}   together imply that $K$ has both  negative and positive eigenvalues. If $(\varphi_W,T)\in \mathcal{D}_4$, Lemmas~\ref{lemma1},~\ref{lemma5}  together imply that $K$ has both  negative and positive eigenvalues. So, in  the case $(\varphi_W,T)\in \mathcal{D}_3\cup \mathcal{D}_4$,  the operator $\Hess$ has both negative and positive eigenvalues and $f_0$ is a saddle point of the objective functional. 

\section{Numerical analysis showing when $f_0=0$ is a global maximum}\label{Sec:5}

\subsection{Reduction to finite-dimensional optimization}
The theoretical analysis above shows that Hessian computed at the control 
$f_0 = 0$ is negative definite. However, this analysis does not say whether control $f_0$ is a global maximum of the objective functional $J_W$, 
or it is a trap for $J_W$, or it is a trap in the weak sense such that  if space of controls is restricted to class of piecewise controls $PConst([0,T],Q) \subset L^2([0,T],\mathbb{R})$, where $Q$ is a compact set, then 
$f_0$ is a local but not global maximum of~$J_W$. Below we  compute the exact analytical form of the objective for piecewise constant controls, and numerically analyze the landscape using GRAPE~\cite{GRAPE} through fminunc function built in MATLAB and also DE, DA through their implementations available in the library {\tt SciPy}. 

Due to the invariance property for $J_W[f]$ with respect to $\theta$ (see 
Lemma~5 in \cite{PechenIl'in2016}), it is sufficient to fix $\theta = 0$. 

\begin{figure}[h!]
	\centering
	\includegraphics[scale=0.3]{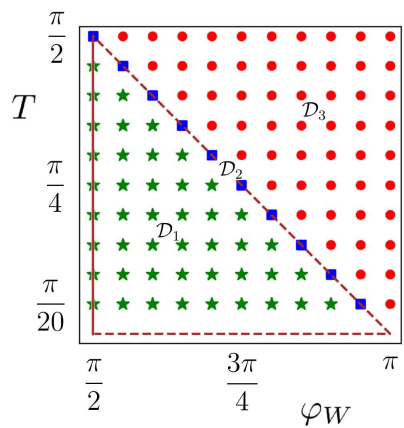} 
	\caption{The grid $\mathcal{G}(\mathcal{D})$: 45~star green markers 
		show the nodes in $\mathcal{D}_1$, 10~square blue markers 
		show the nodes in $\mathcal{D}_2$, and 55~circle red markers 
		show the nodes in $\mathcal{D}_3 \cup 
		\left(\pi, \frac{\pi}{2} \right)$. The triangle boundary of the set $\mathcal{D}_1$ 
		is shown by dotted and solid lines; 
		the points on both dotted lines do not belong to $\mathcal{D}_1$.}
	\label{Section5_figure1}
\end{figure}

In the square domain
\begin{eqnarray*} 
	\hspace*{-1.7cm}
	\mathcal{D} := \mathcal{D}_1 \cup \mathcal{D}_2 \cup \mathcal{D}_3 \cup \left(\pi, \frac{\pi}{2} \right) = \left\{ (\varphi_W, T)~:~ \frac{\pi}{2} \leq \varphi_W \leq \pi, \quad 0 < T \leq \frac{\pi}{2} \right\}, \nonumber
\end{eqnarray*} 
consider the uniform grid  
\begin{eqnarray}
	\label{Section5_f_6}
	\hspace*{-2cm}
	\mathcal{G}(\mathcal{D}) := 
	\Big\{ (\varphi_W^j, T_i) ~:~ \varphi_W^j = \frac{\pi}{2} + \frac{\pi}{20}j,
	\quad j = \overline{0,10}, \quad T_i = \frac{\pi}{20}i, \quad i = \overline{1,10} \Big\}, 
\end{eqnarray}  
which consists of 110 nodes shown on figure~\ref{Section5_figure1}. The grid 
$\mathcal{G}(\mathcal{D}_1)$ (correspondingly, $\mathcal{G}(\mathcal{D}_2)$ and 
$\mathcal{G}(\mathcal{D}_3 \cup \left(\pi, \frac{\pi}{2} \right))$) consists of the nodes of 
$\mathcal{G}(\mathcal{D})$ which belong to $\mathcal{D}_1$ (correspondingly, to 
$\mathcal{D}_2$ and to $\mathcal{D}_3 \cup \left(\pi, \frac{\pi}{2} \right)$). The numerical results below show that in the domains ${\cal D}_1$ and $\mathcal{D}_2$ the control $f_0$ is a global maximum of the objective 
functional $J_W[f]$, in addition to the shown above analytically fact that $J_W[f_0]=1$ for $(\varphi_W, T) \in \mathcal{D}_2$. 

\subsection{Exact form of the objective function for piecewise constant controls}\label{subsec5.2}

For each node $(\varphi_W^j,T_i)$, we restrict the space of controls to the set of piecewise constant controls
\begin{equation}\label{eq:pc}
f(t)=\sum\limits_{k=1}^Ka_k \chi_{[t_k,t_{k+1})}(t),
\end{equation}
where $a_k\in\mathbb R$ is the control amplitude during time interval $[t_k,t_{k+1})$ and $\chi_{[t_k,t_{k+1})}$ is the characteristic function of 
this time interval. Uniform spacing is considered so that $t_{k+1}=k\delta t$, where $\delta t=T_i/K$.  For the final time $T_i$, we take $K=2i$. The minimal value $K=2$ corresponds to the case $T = T_1 = \pi/20$ (see on figure~\ref{Section5_figure1} the bottom horizontal line of the nodes); the maximal value $K =20$ corresponds to the case $T = T_{10} = \pi/2$ (see the upper horizontal line of the nodes). The value $\delta t = \pi/40$ is the same for all $i$.

Control now is vector $\ba=(a_1,\dots,a_K)\in \mathbb R^K$. For GRAPE, no restrictions on $a_k$ are placed. For stochastic methods, the control amplitudes are bounded, so that for some $\nu>0$:
\[
-\nu< a_k< \nu,\qquad k=1,\dots,K.
\]

For piecewise constant controls of the form~(\ref{eq:pc}), the evolution operator is the product of evolution operators at all time intervals, 
\[
U^\ba_T=U_K\dots U_k \dots U_1,
\]
where $U_k=e^{-i(\sigma_z+a_k\sigma_x)\delta t}$. Evolution operator $U_k$ can easly be computed analytically using product rules for Pauli matrices. Indeed, for $A_k=-(\sigma_z+a_k\sigma_x)\delta t$ one has $A_k^2=\delta t^2(1+a_k^2)\cdot \mathbb I_2$. Therefore 
\begin{eqnarray*}
\fl U_k=e^{iA_k}=\sum\limits_{n=0}^\infty \frac{(iA_k)^n}{n!}=\sum\limits_{n=0}^\infty\frac{(iA_k)^{2n}}{{2n}!}+\sum\limits_{n=0}^\infty\frac{(iA_k)^{2n+1}}{{(2n+1)}!}&=&\cos\alpha_k+iA_k\frac{\sin\alpha_k}{\alpha_k}\\
&=&\cos\alpha_k-i\delta t(\sigma_z+a_k\sigma_x)\frac{\sin\alpha_k}{\alpha_k},
\end{eqnarray*}
where $\alpha_k=\delta t\sqrt{1+a_k^2}$. Thus, using sinc function $\sinc\alpha=\sin\alpha/\alpha$, we get
\begin{equation}\label{Eq:U}
\fl U^\ba_T=\Bigl(\cos\alpha_K-i\delta t(\sigma_z+a_K\sigma_x)\sinc\alpha_K\Bigr)\dots \Bigl(\cos\alpha_1-i\delta t(\sigma_z+a_1\sigma_x)\sinc\alpha_1\Bigr).
\end{equation}
This allows to write the objective as a function of the control,
\begin{equation}
\J_W[{\ba}]=\frac{1}{4}|\Tr(W^\dagger U^\ba_T)|^2\to\max\limits_\ba
\end{equation}
(the maximum is taken over $\ba\in\mathbb R^K$ for GRAPE and over $\ba\in[-\nu,\nu]^K$ for DE~and~DA).

\subsection{GRAPE optimization}
GRAPE can be conveniently applied for quantum control problems in cases when gradient of the objective can be analytically computed. Its first advantage is the ability to use analytical expression for the gradient. Second advantage is that GRAPE does not require setting constraints on the amplitude of the control. Third advantage for our control problem is that in the functional space of controls any control other that $f_0$ was proved to be not trap (while restricting the space of controls to piecewise constant controls with fixed $K$ can produce traps). 

For this problem, we compute gradient of the objective
\[
{\rm grad}J_W[\ba]=\frac{1}{2}\Im \left[\Tr (Y^\dagger)\Tr(YV_{k})\right].
\]
Here
\begin{eqnarray*}
Y&=& W^\dagger U_T^\ba,\\
V_{k}&=&U^\dagger_k\dots U^\dagger_1 V U_1\dots U_k.
\end{eqnarray*}
The algorithm starts by randomly generating, with the uniform distribution, the initial control $\ba_0=(a_1,\dots,a_n)$ where each $a_k\in[-A;A]$ ($A=1$ is chosen), and then shifting the control using the computed gradient value. For implementing GRAPE, we apply built in MATLAB function fminunc for unconstrainded optimization using value and gradient of the objective ($-J_W[{\bf a}]$). In numerical simulations, for each node we make runs of GRAPE starting with two random initial points.

\subsection{Differential evolution and dual annealing methods}

The methods of DE and DA are zeroth-order optimization methods which use only values of the objective function and do not use gradient, Hessian, etc. These methods are stochastic methods used for an approximate global optimization. They were applied for example for numerical optimization of coherent and incoherent controls in an open quantum system~\cite{Morzhin_Pechen_PhysParticlesNuclei_2020, Morzhin_Pechen_LJM_2020}. Since the {\tt 
SciPy} library 
gives implementations of these methods to minimize the objective, 
the problem $-J_W[{\bf a}] \to \min\limits_{{\bf a} \in [-\nu, \nu]^K}$ is considered. 
For numerical integration of the dynamical system, the tool {\tt odeint}~\cite{odeint_SciPy} 
available in the {\tt SciPy} library is used. Taking into account the stochastic nature 
of  DE and DA methods, we carry out two runs of each method for each node~$(\varphi_W^j, T_i)$ with a subsequent comparison of the results of all runs.    

\subsection{The numerical simulations for the grid $\mathcal{G}(\mathcal{D}_1)$}

In the domain $\mathcal{D}_1$, for each $l \in Ind(\mathcal{G}(\mathcal{D}_1))$: (1) we numerically find  the optimized vector ${\bf a} = \hat{\bf a}^l$ and the corresponding maximal value $\widehat{J}_W^l$ of the objective function~$J_W[{\bf a}]$ using GRAPE without constraints on the amplitude of the control, as well as the methods of DE and DA for $\nu = 50$; (2)~compare the maximized value of the objective function with the value $J_W^l[0]$.
 
\begin{figure}[h!]
	\includegraphics[scale=0.331]{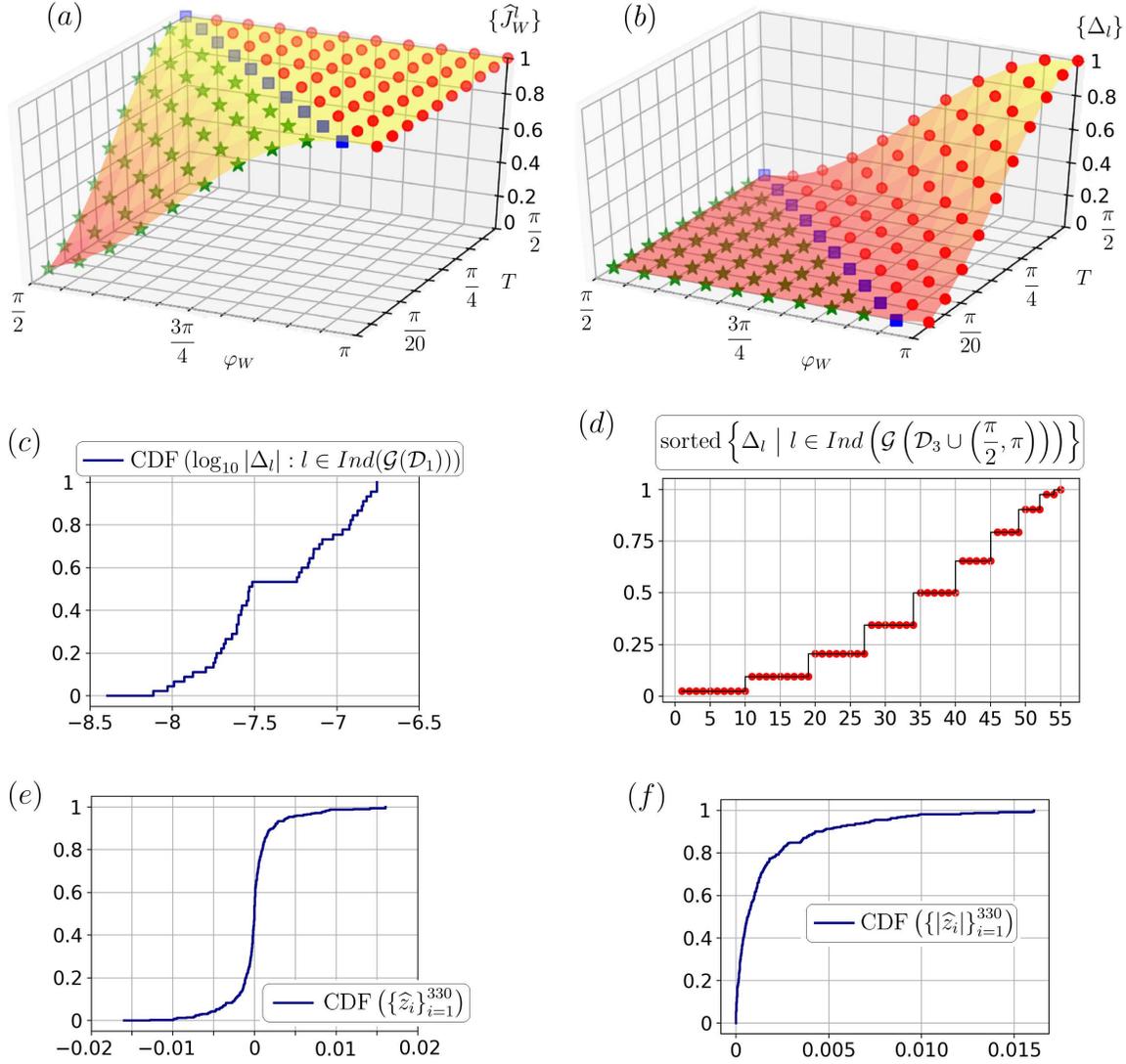} 
	\caption{Subplots (a) and (b) show the table-defined functions $\{\widehat{J}_W^l\}$ and $\{\Delta_l\}$ obtained
		using GRAPE, the methods of DE and DA (for the last two methods $\nu = 
50$ is taken); 
	the corresponding interpolation surfaces are also shown. Here 45~star green markers show points corresponding to the grid $\mathcal{G}(\mathcal{D}_1)$; 10~square blue markers show points corresponding to the grid $\mathcal{G}(\mathcal{D}_2)$; 55~circle red markers show points corresponding to the grid $\mathcal{G}(\mathcal{D}_3 \cup \left(\pi, \frac{\pi}{2} \right))$. Other subplots illustrate the results computed using the methods of DE and DA. Subplot~(c) shows the cumulative distribution function for $\left\{ \log_{10}\left|\Delta_l \right|: l \in Ind(\mathcal{G}(\mathcal{D}_1))  \right\}$. Subplot~(d) shows 55 sorted values $\Delta_l$ for $l \in Ind\left( \mathcal{G}\left( \mathcal{D}_3 \cup \left(\pi, \frac{\pi}{2} 
\right) \right)\right)$. With respect to the grid $\mathcal{G}(\mathcal{D}_1)$, subplots~(e) and~(f) show the cumulative distribution functions for $\left\{ \widehat{z}_i \right\}_{i=1}^{330}$ and $\left\{ \left| \widehat{z}_i \right| \right\}_{i=1}^{330}$, correspondingly.} 
	\label{Section5_figure2}
\end{figure} 

On the grid $\mathcal{G}(\mathcal{D}_1)$, the minimal value among the values $\{\widehat{J}_W^l\}$ is close to 0.0245, and the maximal value is close 0.9755. In other words, exact generation of phase shift gate is not attained for any node of the grid $\mathcal{G}(\mathcal{D}_1)$. On figure~\ref{Section5_figure2}~(a), $45$ star green markers show the values of $J_W$ on this grid. Figure~\ref{Section5_figure2}~(b) shows the graph of the table-defined function   
\begin{eqnarray}\label{Section5_f_11}
\{\Delta_l\} := \{\widehat{J}_W^l - J_W^l[0]\} 
\end{eqnarray} 
corresponding to the grid $\mathcal{G}(\mathcal{D})$. For the grid $\mathcal{G}(\mathcal{D}_1)$, $45$ star green markers at the bottom in figure~\ref{Section5_figure2}~(b) illustrate optimality of the control $f_0$. We have  $\min\limits_{l \in Ind(\mathcal{G}(\mathcal{D}_1))} \left\{ \left|\Delta_l \right| \right\} \approx 4 \cdot 10^{-9}$, $\max\limits_{l \in Ind(\mathcal{G}(\mathcal{D}_1))} \left\{ \left|\Delta_l \right| \right\} \approx 2 \cdot 10^{-7}$, and $\frac{1}{45} \sum\limits_{l \in Ind(\mathcal{G}(\mathcal{D}_1))}  \left|\Delta_l \right| \approx 6 \cdot 10^{-8}$. Figure~\ref{Section5_figure2}~(c) shows the cumulative distribution function for the values
$\left\{ \log_{10}\left|\Delta_l \right|: l \in Ind(\mathcal{G}(\mathcal{D}_1)) \right\}$. The computed values of the function $\{\widehat{J}_W^l\}_{l \in Ind(\mathcal{G}(\mathcal{D}_1))}$ are given in~table~\ref{Section5_table_1}. The computed values of the function $\{\Delta_l\}_{l \in Ind(\mathcal{G}(\mathcal{D}_1))}$ are given in~table~\ref{Section5_table_2}. 
Thus, we conclude that for each node of the grid $\mathcal{G}(\mathcal{D}_1)$ the value $\widehat{J}_W^l$ found by the optimization methods and the value $J_W^l[0]$ computed at the control~$f_0$ coincide with high accuracy.

All 45 numerically optimized controls are represented by $330$ values forming the array
\begin{eqnarray*} 
\widehat{{\bf z}} = \{ \widehat{z}_i \}_{i=1}^{330} := \left\{ 
\widehat{{\bf a}}_k^l:~ k = \overline{1, K_l} , \quad 	
l \in Ind(\mathcal{G}(\mathcal{D}_1))
\right\}. \nonumber
\end{eqnarray*} 
We find that  
$\min\limits_{1 \leq i \leq 330} \left\{ z_i \right\} \approx -0.02$, $\max\limits_{1 \leq i \leq 330} \left\{ z_i \right\} \approx 0.02$, $\min\limits_{1 \leq i \leq 330} \left\{ |z_i| \right\} \approx 5 \cdot 10^{-7}$, and $\frac{1}{330}\sum\limits_{i=1}^{330} |z_i| \approx 0.002$. Figure~\ref{Section5_figure2}~(e,~f) show the cumulative distribution functions for the arrays $\left\{z_i\right\}_{i=1}^{330}$ and $\left\{|z_i|\right\}_{i=1}^{330}$, correspondingly. About 90~\% of the values  $\left\{\widehat{z}_i\right\}_{i=1}^{330}$ are located between $-0.005$ and $0.005$. Thus, we conclude that the all 45 numerically optimized piecewise constant controls are close to the control~$f_0$.  

\subsection{The numerical simulations for the grid $\mathcal{G}(\mathcal{D}_2)$} 
According to~(\ref{JD2}) we have $J_W[0] = 1$ for any $(\varphi_W, T) \in \mathcal{D}_2$. Numerically we also find that with high accuracy $\widehat{J}_W^l \approx 1$ for any $l \in Ind\left(\mathcal{G}(\mathcal{D}_2)\right)$. The values of the function 	$\{\widehat{J}_W^l\}_{l \in Ind(\mathcal{G}(\mathcal{D}_2))}$ are given in table~\ref{Section5_table_1}. On figure~\ref{Section5_figure2}~(b) 10 square blue markers show differences 
$\Delta_l$ for the grid $\mathcal{G}(\mathcal{D}_2)$. The mean value is $\frac{1}{10} \sum\limits_{l \in Ind(\mathcal{G}(\mathcal{D}_2))} \left|\Delta_l \right| \approx 6 \cdot 10^{-6}$. All these ten differences almost 
equal to zero. 

\subsection{The numerical simulations for the grid $\mathcal{G}(\mathcal{D}_3 \cup \left(\pi, \frac{\pi}{2} \right))$} 

Using the numerical optimization methods, we find that for any node of the grid $\mathcal{G}(\mathcal{D}_3 \cup \left(\pi, \frac{\pi}{2} \right))$ 
the maximum value~$\{\widehat{J}_W^l\}$ of the objective functional $J_W$ 
is almost equal to~$1$. This is illustrated by 55 circle red markers in figure~\ref{Section5_figure2}~(a). The values of  $\Delta_l$ for the grid $\mathcal{G}\left( \mathcal{D}_3 \cup \left( \pi, \frac{\pi}{2} \right) \right)$ are shown by 55 circle red markers on figure~\ref{Section5_figure2}~(b). We obtain $\min\limits_{l \in Ind\left( \mathcal{G}\left( \mathcal{D}_3 \cup \left( \pi, \frac{\pi}{2} \right) \right)\right)} \left\{ \left|\Delta_l \right| \right\} \approx 0.0024$, $\max\limits_{l \in Ind\left( \mathcal{G}\left( \mathcal{D}_3 \cup \left(\pi, \frac{\pi}{2} \right) \right)\right)} \left\{ \left|\Delta_l \right| \right\} \approx 1$, and  $\frac{1}{55} \sum\limits_{l \in Ind\left( \mathcal{G}\left( \mathcal{D}_3 \cup \left( \pi, \frac{\pi}{2} \right) \right)\right)} \left|\Delta_l \right| \approx 0.37$.  Figure~\ref{Section5_figure2}~(d) shows sorted values of~$\Delta_l$. The values of $\{\widehat{J}_W^l\}_{l \in Ind\left( \mathcal{G}\left( \mathcal{D}_3 \cup \left( \pi, \frac{\pi}{2} \right) \right) \right)}$ are shown in table~\ref{Section5_table_1}. The values of $\{\Delta_l\}_{l \in Ind\left( \mathcal{G}\left( \mathcal{D}_3 \cup \left( \pi, \frac{\pi}{2} \right) \right) \right)}$ are shown in table~\ref{Section5_table_2}. For each node of~$\mathcal{G}(\mathcal{D}_3 \cup \left(\pi, \frac{\pi}{2} \right))$,  comparing the maximum value~$\widehat{J}_W^l$ 
obtained using the optimization methods with the value $J_W^l[0]$ informs 
that the control~$f_0$ is not globally optimal for this grid. 

All 55 numerically optimized piecewise constant controls are represented by 770 values forming the array 
\begin{eqnarray*}
\widehat{{\bf z}} = \{ \widehat{z}_i \}_{i=1}^{770} := \left\{ \widehat{{\bf a}}_k^l:~ k = \overline{1, K_l},~ l \in Ind\left( \mathcal{G}\left( \mathcal{D}_3 \cup \left( \pi, \frac{\pi}{2} \right) \right)\right) \right\}. \nonumber
\end{eqnarray*}
This array is characterized by  $\min\limits_{1 \leq i \leq 770} \left\{ z_i \right\} \approx -50$, $\max\limits_{1 \leq i \leq 770} \left\{ z_i \right\} \approx 50$, $\min\limits_{1 \leq i \leq 770} \left\{ |z_i| \right\} \approx 0.002$, and $\frac{1}{770}\sum\limits_{i=1}^{770} |z_i| \approx 18$. 

\begin{table}[h!] 
	\caption{\label{Section5_table_1}The obtained values (up to three digits) of the function $\{\widehat{J}_W^l\}_{l=1}^{110}$ for the grid~(\ref{Section5_f_6}). The values are shown on figure~\ref{Section5_figure2}~(a).}  
	\footnotesize
	\begin{tabular}{llllllllllll}
		\br
		\diagbox[width=5em]{$T_i$}{$\varphi_W^j$} & $\varphi_W^1$ & $\varphi_W^2$ & $\varphi_W^3$ & $\varphi_W^4$ & $\varphi_W^5$ & $\varphi_W^6$ & $\varphi_W^7$ & $\varphi_W^8$ & $\varphi_W^9$ & $\varphi_W^{10}$ & $\varphi_W^{11}$ \\ 
		\mr
		$T_{10}$ & 1.0   & 1.0   & 1.0   & 1.0   & 1.0   & 1.0   & 1.0   & 1.0  
 & 1.0   & 1.0 & 1.0 \\  
		$T_{9}$ & 0.976 & 1.0   & 1.0   & 1.0   & 1.0   & 1.0   & 1.0   & 1.0   
& 1.0   & 1.0 & 1.0 \\ 
		$T_{8}$ & 0.905 & 0.976 & 1.0   & 1.0   & 1.0   & 1.0   & 1.0   & 1.0   
& 1.0   & 1.0 & 1.0 \\  
		$T_{7}$ & 0.794 & 0.905 & 0.976 & 1.0   & 1.0   & 1.0   & 1.0   & 1.0   
& 1.0   & 1.0 & 1.0 \\  
		$T_{6}$ & 0.655 & 0.794 & 0.905 & 0.976 & 1.0   & 1.0   & 1.0   & 1.0   
& 1.0   & 1.0 & 1.0 \\  
		$T_{5}$ & 0.5   & 0.655 & 0.794 & 0.905 & 0.976 & 1.0   & 1.0   & 1.0   
& 1.0   & 1.0 & 1.0 \\  
		$T_{4}$ & 0.345 & 0.5   & 0.655 & 0.794 & 0.905 & 0.976 & 1.0   & 1.0   
& 1.0   & 1.0 & 1.0 \\  
		$T_{3}$ & 0.206 & 0.345 & 0.5   & 0.655 & 0.794 & 0.905 & 0.976 & 1.0   
& 1.0   & 1.0 & 1.0 \\  
		$T_{2}$ & 0.095 & 0.206 & 0.345 & 0.5   & 0.655 & 0.794 & 0.905 & 0.976 
& 1.0   & 1.0 & 1.0 \\  
		$T_{1}$ & 0.024 & 0.095 & 0.206 & 0.345 & 0.5   & 0.655 & 0.794 & 0.905 
& 0.976 & 1.0 & 1.0 \\ 
		\br 
	\end{tabular} 
	\normalsize
\end{table}

\begin{table}[h!] 
	\caption{\label{Section5_table_2}The obtained values (up to three digits) of the function $\{\Delta_l\}_{l=1}^{110}$ for the grid~(\ref{Section5_f_6}). The values are shown on figure~\ref{Section5_figure2}~(b).}  
	\footnotesize
	\begin{tabular}{llllllllllll}
		\br
		\diagbox[width=5em]{$T_i$}{$\varphi_W^j$} & $\varphi_W^1$ & $\varphi_W^2$ & $\varphi_W^3$ & $\varphi_W^4$ & $\varphi_W^5$ & $\varphi_W^6$ & $\varphi_W^7$ & $\varphi_W^8$ & $\varphi_W^9$ & $\varphi_W^{10}$ & $\varphi_W^{11}$ \\ 
		\mr
		$T_{10}$ & 0 & 0.024 & 0.095 & 0.206 & 0.345 & 0.5   & 0.655 & 0.794 & 0.905 & 0.976 & 1.0   \\ 
		$T_{9}$ & 0 & 0     & 0.024 & 0.095 & 0.206 & 0.345 & 0.5   & 0.655 & 0.794 & 0.905 & 0.976 \\  
		$T_{8}$ & 0 & 0     & 0     & 0.024 & 0.095 & 0.206 & 0.345 & 0.5   & 0.655 & 0.794 & 0.905 \\  
		$T_{7}$ & 0 & 0     & 0     & 0     & 0.024 & 0.095 & 0.206 & 0.345 & 0.5   & 0.655 & 0.794 \\   
		$T_{6}$ & 0 & 0     & 0     & 0     & 0     & 0.024 & 0.095 & 0.206 & 0.345 & 0.5   & 0.655 \\   
		$T_{5}$ & 0 & 0     & 0     & 0     & 0     & 0     & 0.024 & 0.095 & 0.206 & 0.345 & 0.5   \\   
		$T_{4}$ & 0 & 0     & 0     & 0     & 0     & 0     & 0     & 0.024 & 0.095 & 0.206 & 0.345 \\   
		$T_{3}$ & 0 & 0     & 0     & 0     & 0     & 0     & 0     & 0     & 0.024 & 0.095 & 0.206 \\   
		$T_{2}$ & 0 & 0     & 0     & 0     & 0     & 0     & 0     & 0     & 0 
    & 0.024 & 0.095 \\  
		$T_{1}$ & 0 & 0     & 0     & 0     & 0     & 0     & 0     & 0     & 0 
    & 0     & 0.024 \\  
		\br 
	\end{tabular} 
	\normalsize
\end{table} 

\subsection{Minimal final times}

As~figure~\ref{Section5_figure2}~(a) shows, if we fix the value $\varphi_W^j$ and look for a minimal time to reach the boundary value $J_W = 1$, 
then we can take  
\begin{eqnarray*} 
	T_i = \pi - \varphi_W^j, \quad j \in \overline{0,10}, \quad i \in \overline{1,10}, \quad 
	f(t) = 0 \quad \forall t \in [0, T_i],
\end{eqnarray*} 
i.e. we can take the corresponding node of the grid $\mathcal{G}(\mathcal{D}_2)$ and the control $f_0$. 

\subsection{The piecewise constant controls with two variables}

In this subsection we consider the case $K = 2$ and  $T_1 = {\pi}/{20}$, so that $\delta T=T_1/K=T/2$. This case is considered as it allows for a simple visualization of the typical structure of the control landscape, similarly to that done in~\cite{Pechen2012,Larocca2018} for a different objective functional. The objective functional $J_W[f]$ in this case reduces to the objective function $J_W[{\bf a}]$ of ${\bf a} = (a_1, a_2)\in\mathbb R^2$. 

With the notations $\alpha_i=\delta t\sqrt{1+a_i^2}$ for $i=1,2$, the 
objective function takes the exact analytical form
\begin{eqnarray}
J_W[{\bf a}]&=&2\cos\phi_W\Bigl\{\cos\alpha_1\cos\alpha_2-\delta t^2(1+a_1a_2)\sinc\alpha_1\sinc\alpha_2\Bigr\}\nonumber\\
&&-2\delta t\sin\phi_W\Bigl\{\sinc\alpha_1\cos\alpha_2+\sinc\alpha_2\cos\alpha_1\Bigr\}.
\end{eqnarray}

\begin{figure}[h!]
	\centering
	\includegraphics[width=1\textwidth]{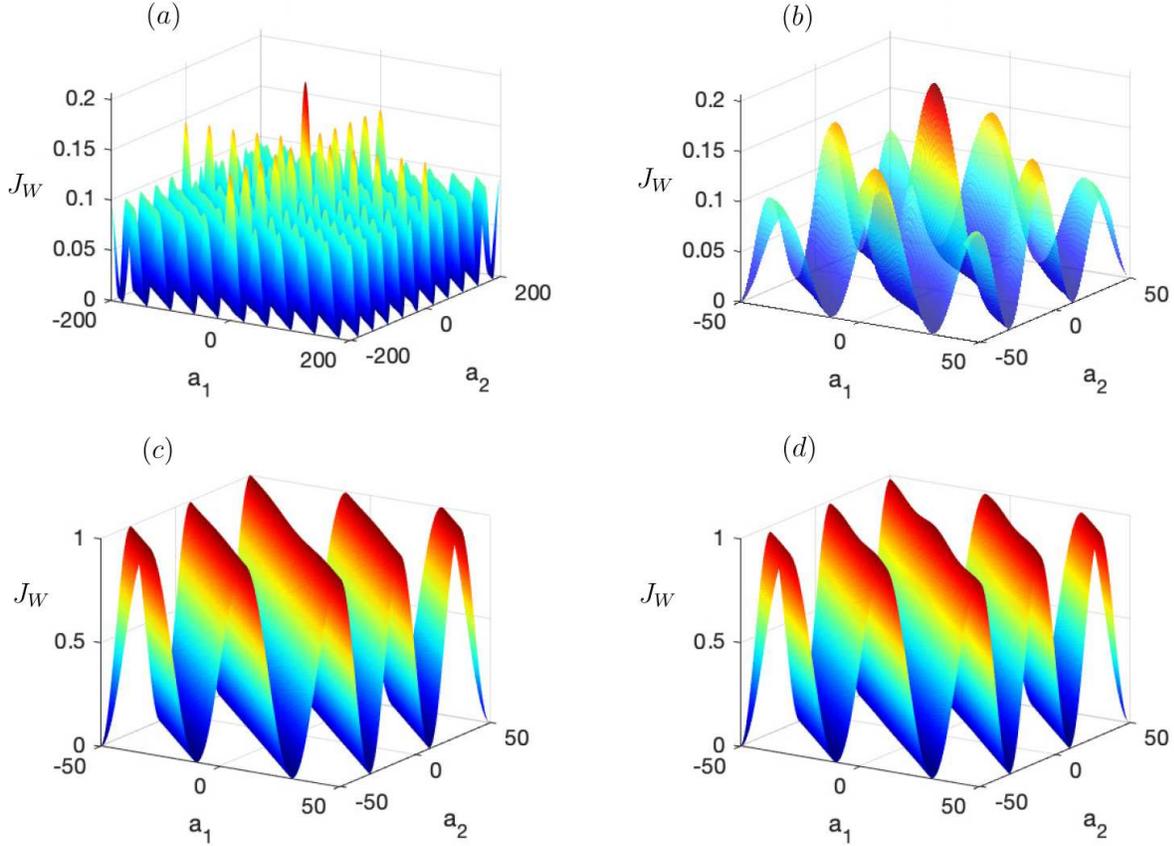}  
	\caption{The dynamical control landscapes for $T_1 = {\pi}/{20}$ and $K = 2$: (a),~(b)~for $\left(\varphi_W, T_1 \right) = \left(\frac{3\pi}{5}, \frac{\pi}{20} \right) \in \mathcal{G}(\mathcal{D}_1)$; (c)~for $\left(\varphi_W, T_1 \right) = \left(\pi, \frac{\pi}{20} \right) \in \mathcal{G}(\mathcal{D}_3)$; (d)~for $\left(\varphi_W, T_1 \right) = \left(\frac{19\pi}{20}, \frac{\pi}{20} \right) \in \mathcal{G}(\mathcal{D}_2)$.}
	\label{Section5_figure3}
\end{figure}

Define on the set $[-\nu, \nu] \times [-\nu, \nu]$ the uniform grid with the step $\Delta a = 1$,
\begin{eqnarray*} 
	\hspace*{-2.5cm}
	\left\{ \left(a_1^s, a_2^q \right): ~ a_1^s = -\nu + s \Delta a, \quad 
s = \overline{0,N}, 
	\quad a_2^q = -\nu + q \Delta a; 
	\quad q =\overline{0,N}, \quad \Delta a = \frac{2\nu}{N} \right\}\, .
\end{eqnarray*} 
For each node $(a_1^s, a_2^q)$, the corresponding control is denoted as $f^{s,q}$. Consider the cases $\left( \varphi_W, T_1 \right) = \left(\frac{3\pi}{5}, \frac{\pi}{20} \right) \in \mathcal{G}(\mathcal{D}_1)$,     $\left(\varphi_W, T_1 \right) = \left(\frac{19\pi}{20}, \frac{\pi}{20} \right) \in \mathcal{G}(\mathcal{D}_2)$, and $\left(\varphi_W, T_1 \right) = \left(\pi, \frac{\pi}{20} \right) \in \mathcal{G}(\mathcal{D}_3)$. For each case, the corresponding control landscape is plotted on figure~\ref{Section5_figure3} by computing the values of the objective function $J_W[{\bf a}]$ for each node $(a_1^s, a_2^q)$. The presence of traps in these control landscapes is due to the strong constraints posed by the form 
of the control.  Subplots~(a) and~(b) for the node $\left( \varphi_W, T_1 
\right) = \left(\frac{3\pi}{5}, \frac{\pi}{20} \right)\in\mathcal{D}_1$ 
show that: (1)~ ${\bf a} = (0,0)$ is a global maximum with the objective value significantly smaller than~1; (2)~traps exist. Subplot~(c) for the node $\left(\varphi_W, T_1 \right) = \left(\pi, \frac{\pi}{20} \right) \in \mathcal{D}_3$, shows that ${\bf a} = (0,0)$ is a saddle point.  In this case, the maximum value $J_W[{\bf a}]=1$ is reachable (in~table~\ref{Section5_table_1}, see for the case $(\varphi_W^{11}, T_1)$), but not at ${\bf a} = (0,0)$. Subplot~(d) for the node $\left(\varphi_W, T_1 
\right) = \left(\frac{19\pi}{20}, \frac{\pi}{20} \right) \in \mathcal{D}_2$ shows that  the global maximum with objective value~1 is achieved at 
${\bf a} = (0,0)$ implying precise generation of the quantum gate by the control~$f_0$. 

\section{Conclusion} 

In this work, we studied the problem of ultrafast controlled generation of single-qubit phase shift quantum gates. This problem can be reduced to the problem of maximization  of the objective functional $J_W$ parameterized by $\varphi_W\in(0,\pi]$ in the case of the fixed time  $T <T_0=\pi/2$. In~\cite{PechenIl'in2016} it was proved that  the only possible trap 
for this objective functional is the special control $f_{0}=0$ and it could potentially be trap only if $\varphi_W \in [\frac\pi 2,\pi]$ and $T<\pi-\varphi_W$.  It was  known that this special control under these conditions is a critical point of the objective functional, but it was not known whether it is a saddle point, a global extremum point, or a trap.  In 
this paper, we have investigated the Hessian of the objective functional $J_W$ at $f_0$ for various values $\varphi_W \in (0,\pi]$  and for small times $T$. In this case, the Hessian is an integral self-adjoint operator. Considering the inverse differential operator to the Hessian, we investigated its spectrum and eigenvalues. We show that for $\varphi_W \in [\frac\pi 2,\pi] $ and $T\in(\pi-\varphi_W,\frac{\pi}2) $ and for $\varphi_W\in (0,\frac\pi 2)$  and $T\in (0,\frac\pi 2]$ such that 
$\varphi_W+T\neq \frac{\pi}2$ the special control $f_0 = 0$ is a saddle 
point of the objective functional $J_W$. This result was previously obtained in~\cite{PechenIl'in2016} by another method. A new result of this paper is the proof that for $\varphi_W \in [\frac \pi 2,\pi]$ and $T <\pi-\varphi_W$ the Hessian is a negative definite operator. Thus, it is rigorously proved that in this case $f_0$ is either a global maximum point, a trap, or a trap in the weak sense. The numerical analysis is further performed to show that this control is a global maximum point. The numerical results also show that for $\frac{\pi}{2} \leq \varphi_W \leq \pi$ and $0 < 
T \leq \frac{\pi}{2}$ achieving the objective functional value~1, i.e., providing exact generation of phase shift gate, requires a final time $T$ being not less than the minimal time $T_{\min} = \pi - \varphi_W$. The exploited method based on Hessian analysis is potentially quite general and could be applied to the analysis of control landscapes for multi-level 
quantum systems and other control problems such as maximizing the transition probability to a target state and optimizing average value of a target observable, while the exact analysis may often be problem specific.

\ack This work was funded by Russian Federation represented by the Ministry of Science and Higher Education (grant number 075-15-2020-788).

\end{document}